# Manufacturing high-Q superconducting α-tantalum resonators on silicon wafers

D. P. Lozano[1], M. Mongillo[1], X. Piao[1], S. Couet[1], D. Wan[1], Y. Canvel[1], A. M. Vadiraj[1], Ts. Ivanov[1], J. Verjauw[1,2], R. Acharya[1,3], J. Van Damme[1,3], F. A. Mohiyaddin[1], J. Jussot[1], P. P. Gowda[1], A. Pacco[1], B. Raes[4], J. Van de Vondel[4], I. P. Radu[1], B. Govoreanu[1] J. Swerts[1], A. Potočnik[1], and K. De Greve[1,3]

[1]Imec, Kapeldreef 75, Leuven, B-3001, Belgium,
[2]Department of Materials Engineering (MTM), KU Leuven, Leuven, B-3000, Belgium
[3]Department of Electrical Engineering (ESAT), KU Leuven, Leuven, B-3000, Belgium
[4]Departmentof Physics and Astronomy, KU Leuven, Leuven, B-3000, Belgium

**Abstract**
The performance of state-of-the-art superconducting quantum devices is currently limited by microwave dielectric losses at different surfaces and interfaces. α-tantalum is a superconductor that has proven effective in reducing dielectric loss and improving device performance due to its thin low-loss oxide. However, without the use of a seed layer, this tantalum phase has so far only been realised on sapphire substrates, which is incompatible with advanced processing in industry-scale fabrication facilities. Here, we demonstrate the fabrication of high-quality factor α-tantalum resonators directly on silicon wafers over a variety of metal deposition conditions and perform a comprehensive material and electrical characterization study. By comparing experiments with simulated resonator loss, we demonstrate that two-level-system loss is dominated by surface oxide contributions and not the substrate-metal interface. Our study paves the way to large scale manufacturing of low-loss superconducting circuits and to materials-driven advancements in superconducting circuit performance.

**INTRODUCTION:**
Increasing qubit performance is essential for enhancing the capabilities of Noisy Intermediate-Scale Quantum (NISQ) processors and crucial for reducing the error-correction overhead in future fault-tolerant quantum computers. Historically, several approaches have been explored to increase their performance such as, participation ratio engineering[1] (typically resulting in bigger qubits), optimal control[2,3], shielding[4] and signal filtering[5]. In contrast, the number of advancements based on understanding microscopic material properties – including the judicious increase of the materials toolbox - is rather limited, with most published work focusing on only a few well documented materials such as Al[6,7,8,9,10,11], Nb[12,13,14,15], and TiN[16,17].

Only recently, the suite of materials for superconducting quantum technology was significantly expanded, markedly resulting in qubit relaxation times as high as 0.5 ms for 2D transmon qubits by using α-tantalum (α-Ta)[18,19]. It has been suggested that the simpler native oxide structure of tantalum[20] compared to native oxides structures from other materials such as niobium (Nb)[21] leads to fewer microwave loss. Furthermore, these studies employed sapphire as substrate, which has lower dielectric loss[22,23,24] than the high-resistivity silicon (Si)[25,26] substrate typically used in superconducting circuits. However, the use of sapphire substrate is incompatible with fabrication processes and requirements needed to manufacture and integrate large numbers of qubits in 300 mm industry-scale facilities.

Fabricating superconducting circuits that use α-tantalum as superconducting film on Si is possible with the help of seed layers such as Nb[27], tantalum nitride (TaN)[28] and titanium nitride (TiN)[29]. However, this method increases the processing complexity and implies the creation of an extra interface that can host two-level system[30] (TLS) defects, thus potentially degrading the performance of the superconducting device.

In this study, we demonstrate for the first time the fabrication high-Q factor α-Ta coplanar-waveguide resonators directly on Si without the need of a seed layer and establish a wide temperature processing window that allows the growth of this Ta phase. We characterize the microwave performance of the devices and show that our results are consistent with TLS-induced dissipation. We further implement and perform in-depth characterization of the metal-air (M-A), metal-substrate (M-S), and substrate-air (S-A) interfaces by combining spectroscopy and microscopy techniques to locate loss sources and demonstrate that surface oxides are the main TLS loss contributors in our samples. These findings highlight the significance of material development in enabling large-scale manufacture of superconducting circuits, as well as material characterisation in identifying dominant loss causes in order to improve the performance of superconducting devices.

## RESULTS:
### Characterization of alpha tantalum films

We deposit α-Ta films on Si substrates at 400°C[31] with a nominal thickness of 100 nm. To verify the growth of the correct phase, we measure the X-diffraction (XRD) spectrum of the film [Figure 1(a)]. Two diffraction peaks corresponding to α-Ta are visible, indicating the polycrystalline nature of our films, while no peaks corresponding to β-Ta are present. The microstructure and crystallinity of α-Ta highly depends on the substrate of choice[32,33], therefore, we further characterize the microstructure of the film by using scanning transmission electron microscopy (STEM). Figure 1(b) depicts a cross sectional view of an as-deposited tantalum film, confirming it has a polycrystalline structure and the presence of a native oxide layer about 2.8 nm thick. It is known that impurities in the bulk of some superconducting films like Nb[34] and Al[35] affect their superconducting properties. Therefore, to investigate the impurity concentration in our Ta film, we use Time of Flight Secondary Ion Mass Spectrometry (ToF-SIMS) [Figure 1(c)]. Oxygen (O), hydrogen (H), fluorine (F), carbon (C) and chlorine (Cl) are present within the bulk of the film. The signal detected for H⁻, F⁻, C⁻ and Cl⁻ is just above the detection level, indicating that the amount of these elements in the bulk of the film is negligible. A Cl⁻ signal increase can be observed when approaching the tantalum-silicon interface, but this increase is an artifact, a result of Cl⁻ high negative ionization probability and a matrix effect (going from a Ta matrix to a Si matrix). Moreover, an oxide layer is present on the sample together with C, Cl, H, F contaminants. The signal corresponding to contaminants decays faster than the one associated with Ta and O, indicating that contaminants lie on top of the oxide layer. We also performed electrical characterization of the film by measuring electrical resistivity and critical temperature ($T_c$). While room-temperature resistivity (14.66 μΩcm) is comparable to the typical resistivity of α-Ta films with similar thicknesses[36], the superconducting transition temperature [$T_c$ ~2.9 K, Figure 1(d)] is lower than the $T_c$ expected for a 100 nm α-Ta[37]. Nevertheless, sputtered α-Ta with $T_c$ similar to ours have been reported before[38]. Finally, we find that α-Ta grows also at 450°C and 500°C deposition temperatures with comparable morphological and electrical properties as the 400°C film (supplementary material).

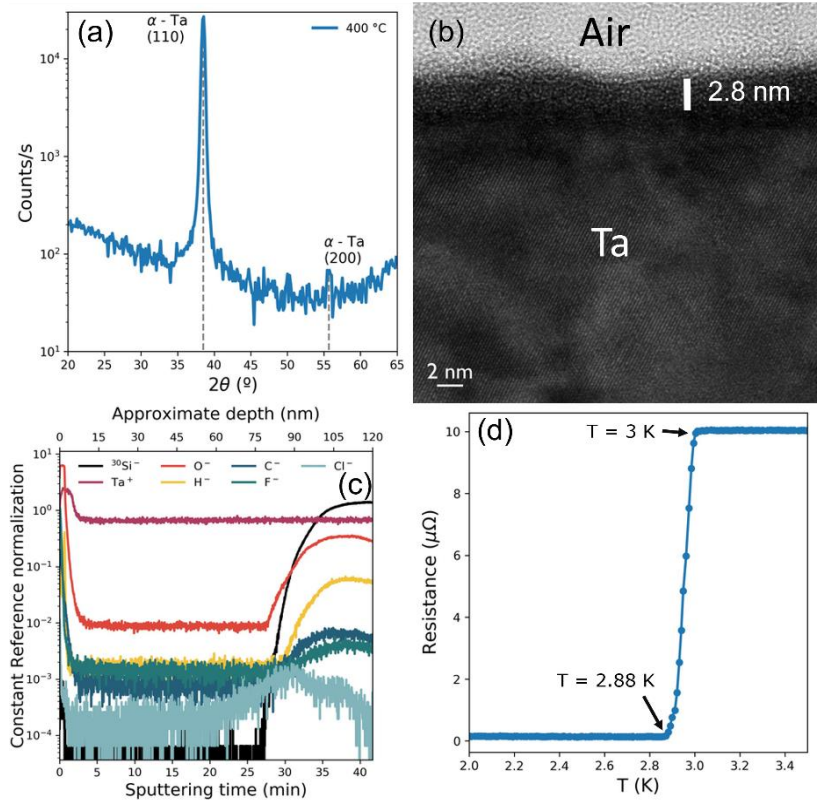

Figure 1. Ta film characterization. (a) XRD spectrum. (b) Bright-field STEM cross section of the film. (c) Film TOF-SIM spectra for $^{30}Si^-$, $Ta^+$, $O^-$, $H^-$, $C^-$, $F^-$ and $Cl^-$. (d) Superconducting transition of the Ta film observed around 2.9 K.

Next, we pattern standard coplanar-waveguide resonators using the 400°C, 450°C and 500°C α-Ta films to study microwave loss sources and potential loss differences between the three film types. Resonator fabrication is performed in a 300 mm fabrication facility at IMEC using industry-standard fabrication process[39]. To study the morphology and composition of the M-A, M-S and S-A interfaces, regions that typically host dominant loss sources[40], we conduct a detailed characterization of patterned resonators using STEM and Energy-Dispersive X-ray Spectroscopy (EDS). Figure 2(a) shows an overview of the different interfaces studied corresponding to a resonator patterned using the α-Ta film deposited at 400°C. The top M-A interface [Figure 2(b)] is covered by a 3.7 nm tantalum oxide layer, while the tantalum oxide at the side wall is about 6 nm thick. The oxide thickness on patterned Ta is larger compared to the 2.8 nm oxide in the blanket Ta layer because of the oxygen plasma[41] treatments used during the fabrication flow. This can also explain the difference in oxide thickness between the top and the sidewall of the patterned Ta film. During the oxygen plasma strip used to remove the resist mask the sidewall is exposed longer to the oxygen plasma than the top surface, which was initially protected by a $SiO_2$ hard mask (see Methods). The S-A interface shown [Figure 2(c)] contains approximately 2.5 nm thick $SiO_2$ layer, while M-S interface [Figure 2(d)] is approximately 4-5 nm thick. The M-S layer consists of a mixture of Si and Ta, as confirmed by the contrast change in the high-angle annular dark-field-STEM image of the same region (Figure 2(e)) and occurs due to the short-range diffusion of Si into Ta sputtered layers at elevated temperature[42,43]. Furthermore, Figure 2(f) shows the oxygen EDS map of the same region as in Figure 2(d). The absence of a marked signal across the Si-Ta interface indicates that this interface does not contain notable oxygen contamination, implying negligible oxide growth after substrate clean and before metal deposition. The blue dots visible correspond to the background signal of the detector. Similar results are obtained also for 450°C and 500°C patterned Ta films (see supplementary information).

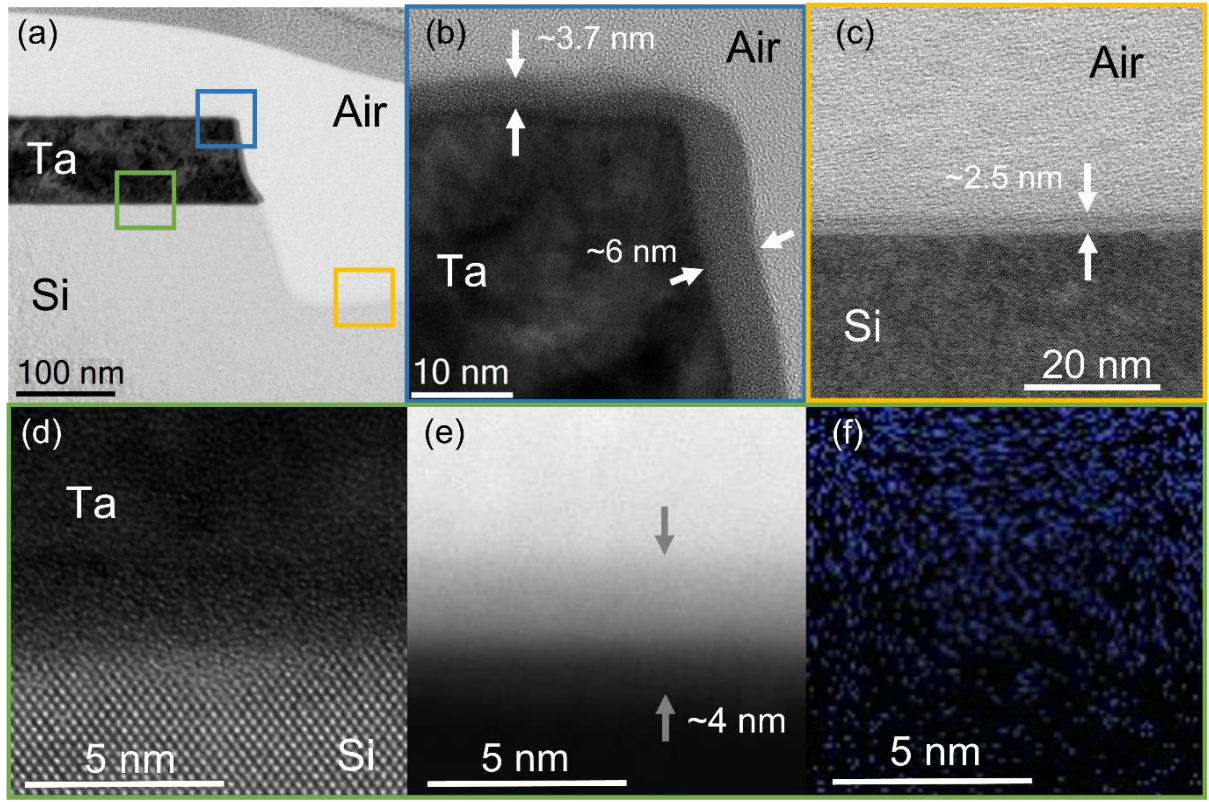

Figure 2 STEM and EDS images of a 400°C α-Ta resonator cross section. (a) low magnification STEM cross section of Ta resonator. The coloured squares mark the areas where high magnification images were taken. (b) STEM cross section of the metal-air interface. Arrows indicate the tantalum oxide layer. (c) Annular bright-field-STEM image of the substrate-air interface. The arrows indicate the silicon oxide layer present. (d) STEM cross section of the substrate-metal interface. (e) High-angle annular dark-field-STEM cross section of the substrate-metal interface. The arrows indicate the extension of the Si-Ta interfacial layer. (f) Cross section oxygen EDS map of the substrate-metal interface.

**Resonator measurement:**

We characterize the performance of the superconducting coplanar waveguide resonators by performing $S_{21}$ transmission spectrum measurements at 10 mK. All eight resonators on a chip have 4.5 μm large gap and 10 μm wide central trace, while their frequencies are distributed between 4 and 8 GHz. A fitting routine[44] is used to extract the resonant frequency $f_r$, internal quality factor $Q_i$, and coupling quality factor $Q_c$ as a function of the mean photon number in the different chips investigated. Figure 3(a) shows $Q_i$ of measured resonators fabricated using the 450°C Ta film from two different chips as a function of microwave power expressed with mean photon number $\langle n \rangle$. The first chip labelled as "reference" corresponds to as fabricated chip without any post-processing, and the second chip labelled as "HF-treated" corresponds to a chip that received a post-processing hydrofluoric acid clean (HF-clean) step not longer than 12h before the cooldown of the dilution refrigerator. Apart from chip mounting and wire bonding, HF-treated chips were stored in the $N_2$ atmosphere during the 12h waiting period. HF post-processing treatment is employed to reduce surface oxides, which are the major source of loss in superconducting quantum devices[45]. In this work we used the same treatment applied to Nb lumped element resonators (60s in 10 vol% HF) used by Verjauw[9] since it has proven effective to completely remove $SiO_2$, passivate the Si surface, and remove surface niobium oxides, resulting in a $Q_i$ increase by a factor of seven.

In both reference and HF-treated chips, the observed $Q_i$ power dependence indicates the presence of dominating TLS depolarisation in our devices[46]. In total, twelve different α-Ta chips (six reference chips and six HF-treated chips) fabricated using the 400°C, 450°C and 500°C α-Ta films were measured in two different sample holders: sample holder A and B. Sample holder B has been optimized to reduce the residual magnetic fields that may affect resonator performance. The six reference chips have mean low power quality factor ($Q_{i,\,low}$) values [Figure 4(b)] ranging between $0.4 \times 10^6$ and $0.8 \times 10^6$, with chips measured in sample holder B having higher mean $Q_{i,\,low}$ due to the reduction in magnetic-field induced loss. The HF-treated chips have $Q_{i,\,low}$ ranging between $1.0 \times 10^6$ and $2.5 \times 10^6$. Similarly, chips measured in sample holder B have higher mean $Q_{i,\,low}$, although with a significantly larger spread. In addition, there are no significant differences between the chips fabricated using the 400°C, 450°C and 500°C, which agrees with comparable morphological structure of the three different films discussed above. Moreover, the observed increase in $Q_{i,\,low}$ in the HF-treated chip compared to the reference chips is due to the HF oxide removal in both the M-A and the S-A interfaces.

At high photon numbers $Q_{i,\,high}$ [Figure 4(c)], the reference chips measured using sample holder A display mean $Q_{i,\,high}$ values ranging between $0.8 \times 10^6$ and $2.0 \times 10^6$, while the chips measured in sample holder B have mean $Q_{i,\,high}$ values ranging between $24 \times 10^6$ and $28 \times 10^6$. In the case of the HF-treated chips measured using sample holder A mean $Q_{i,\,high}$ values vary between $1.6 \times 10^6$ and $1.8 \times 10^6$, while mean $Q_{i,\,high}$ values of HF-treated chips in sample holder B vary between $18 \times 10^6$ and $34 \times 10^6$. This difference of mean $Q_i$ values between sample holder A and B is due the improvement in the magnetic shielding provided by the sample holder B. We note that there is no significant difference in $Q_{i,\,high}$ between reference and HF-treaded chips.

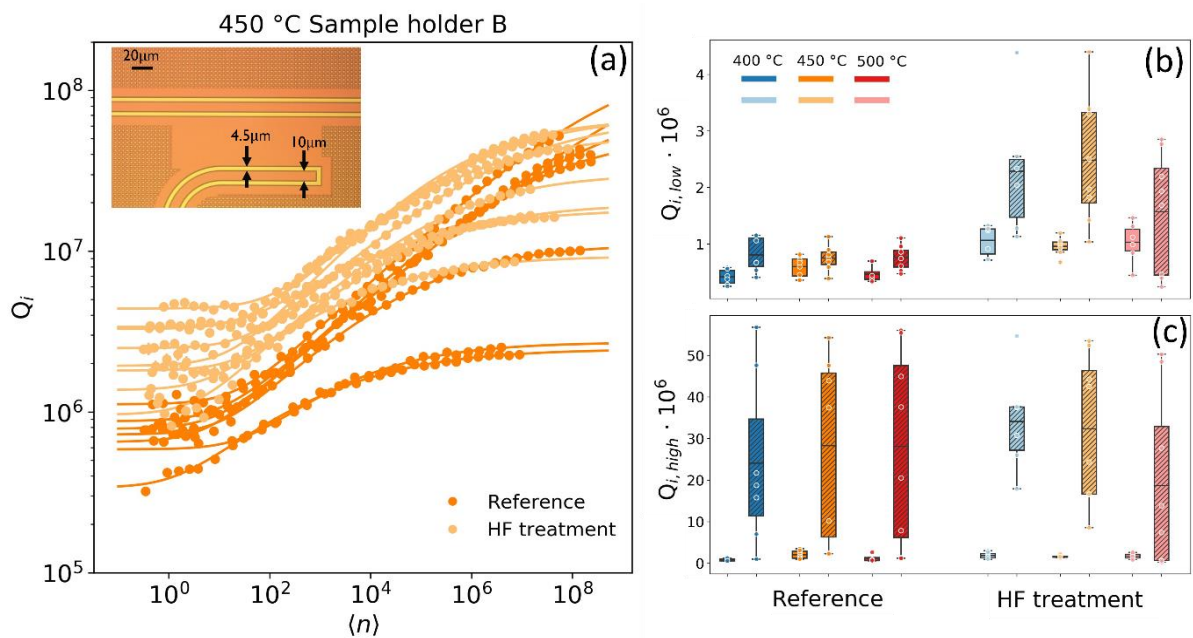

Figure 3 Resonator measurement summary. (a) Internal quality factor as a function of mean photon number measured in sample holder B for the reference and HF-treated Ta resonators fabricated using the 450°C α-Ta film. The lines are fittings to equation 1. The inset is a micrograph of one of the measured resonators. (b-c) Boxplot summary of $Q_{i,\,low}$ ($Q_{i,\,high}$), of the reference and HF-treated Ta resonators chips. The boxes extend from the first quartile to the third quartile of $Q_{i,\,low}$ ($Q_{i,\,high}$), with a line at the mean. The whiskers extend from the box by 1.5 times the inter-quartile range. Dots indicate experimental measurements. Boxes with no texture correspond to measurements taken using sample holder A, while boxes with a hatched pattern correspond to measurements taken using sample holder B.

## Discussion:

To isolate material-specific TLS defect microwave loss we analyse $Q_i$ as a function of $\langle n \rangle$ using a TLS-based loss model described by equation 1 [see Figure 3(a)]

$$\frac{1}{Q_i} = F\tan\delta_0 \frac{\tanh\left(\frac{hf_r}{2k_BT}\right)}{\left(1+\frac{\langle n \rangle}{n_c}\right)^b} + \delta_{\text{other}} \qquad (1)$$

where F is the participation ratio, $\tan\delta_0$ is the intrinsic loss tangent for the material containing the TLS and $\delta_{\text{other}}$ is the contribution from power independent non-TLS loss. Similarly, $n_c$ is the critical photon number equivalent to the saturation field of different TLSs and b is a phenomenological parameter, which is 0.5 for non-interacting TLS defects[47] and lower than 0.5 in the presence of TLS-TLS interactions[48].

The data agrees well with the above model for all measured resonators [Figure 3(a) and Supplementary Figures S4(a-f)]. The *b* parameter is generally less than 0.5 [Supplementary Figure S5(a)], which aggress with previously b values reported[21] in resonators with $Q_i$ around and above 1x10$^6$ and indicating that TLS-TLS interactions are present. Critical photon numbers in α-Ta devices are found to be $n_c$ = 1-50, except for the two HF-treated chips fabricated with 400°C α-Ta where $n_c$ can reach up to 650 photons [Supplementary Figure S5(b)]. The TLS loss expressed by $F\tan\delta_0$ for the reference chips measured in both sample holders A and B is similar Figure 4(a)]. This is expected since this type of loss is mostly related to surface oxides and defects at other interfaces that do not depend on the sample environment. Their weighted mean $F\tan\delta_0$ values range between 1.04x10$^{-6}$ and 1.14x10$^{-6}$. The weighted mean $F\tan\delta_0$ is smaller in the HF-treated chips compared to the reference ones, and it varies between 0.35x10$^{-6}$ and 0.45x10$^{-6}$. These results are summarized in Table 1.

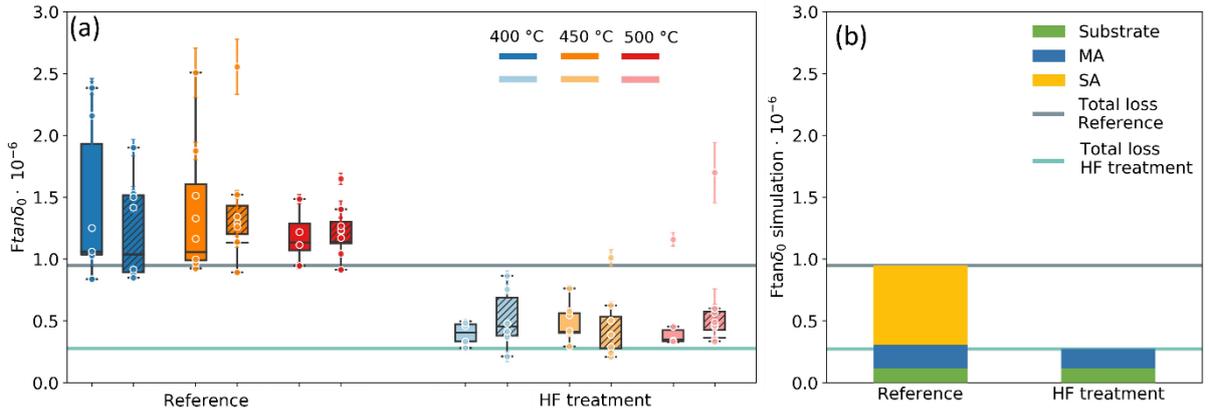

Figure 4 TLS loss measurement and simulation results. (a) Boxplot of the $F\tan\delta_0$ extracted from fitting the reference and HF-treated Ta resonators resonator data to equation 1. The boxes extend from the first quartile to the third quartile of $F\tan\delta_0$ with a line at the mean. The whiskers extend from the box by 1.5 times the inter-quartile range. Dots indicate individual $F\tan\delta_0$ values extracted from fitting. Boxes with no texture correspond to measurements taken using sample holder A while boxes with a hatched pattern corresponds to measurements taken using sample holder B. The horizontal lines indicate the total simulated $F\tan\delta_0$ values. (b) Bar plot representing the mean values across the different chips of simulated $F\tan\delta_0$ for the substrate, metal-air and substrate-air interfaces for the reference and HF-treated chips. The horizontal lines indicate the total simulated $F\tan\delta_0$, the same as in panel (a).

Table 1 Weighted mean values for $F\tan\delta_0$ extracted from resonator fitting in the reference and HF-treated chips together with the simulated values.

|  | Weighted mean $F\tan\delta_0$ (x $10^{-6}$) Sample holder A | | Weighted mean $F\tan\delta_0$ (x $10^{-6}$) Sample holder B | | Simulated $F\tan\delta_0$ (x$10^{-6}$) | |
|---|---|---|---|---|---|---|
|  | Reference | HF treatment | Reference | HF treatment | Reference | HF treatment |
| **400 °C** | 1.06 ± 0.48 | 0.40 ± 0.09 | 1.04 ± 0.34 | 0.44 ± 0.15 | 0.93 | 0.27 |
| **450 °C** | 1.06 ± 0.23 | 0.40 ± 0.14 | 1.13 ± 0.30 | 0.28 ± 0.10 | 0.93 | 0.27 |
| **500 °C** | 1.14 ± 0.21 | 0.35 ± 0.09 | 1.13 ± 0.20 | 0.36 ± 0.07 | 0.98 | 0.29 |

This TLS loss reduction can be explained by a reduction in loss associated with the M-A and S-A. The HF treatment employed is sufficient to completely remove the $SiO_2$ from the S-A interface and passivate the Si surface to avoid oxide regrow before the chips are cooled down and measured. This HF treatment also reduces the tantalum oxide thickness by 15%-18% as indicated by X-ray photoemission spectroscopy (XPS) measurements (Supplementary Figure S6).

To further understand the contribution of each interface to the total TLS loss in our devices we simulate interface loss contributions of both reference and HF-treated chips [Figure 4(b)]. Loss contributions are calculated using interface participation ratios from finite-element electrostatic simulations based on resonator profile from the STEM images. Dielectric loss tangents for Si[40], $SiO_2$[40] and $Ta_2O_5$[49,50] are extracted from literature. The TLS loss contribution from $TaO_x$ is neglected due to its small percentage compared the total tantalum oxide amount as shown by the XPS data (supplementary information). In reference chips, the S-A interface accounts for around 68% of TLS loss, whereas the M-A interface accounts for 20% and the Si substrate for 12%. In the case of the HF-treated chips, around 57% of the TLS loss is due to the M-A interface, while 43% is due to the Si substrate. The horizontal lines in Figures 4(a-b) represent the total TLS loss from the different simulations performed. We note that the simulated $F\tan\delta_0$ reasonably agree with the measured $F\tan\delta_0$ values, however, a systematic underestimation can be observed. This might be because the $Ta_2O_5$ loss tangent value used in our calculations is obtained at room temperature, while our measurements occur at 10 mK, or due to an additional loss contribution from the S-M interface, which was not included in our simulations as we are experimentally not able to distinguish between bulk Si and M-S losses.

Although no oxide is detected at the S-M interface, as shown in the oxygen EDS map in Figure 2(f), there is evidence of Si diffusion into the Ta film. Different tantalum silicide compounds are known to form at the Ta-Si interface when exposed to elevated temperatures. The formation of pentatantalum trisilicide ($Ta_5Si_3$) occurs when Ta films deposited on Si substrates are annealed at 550°C – 600°C for one hour[51], while tantalum silicide ($TaSi_2$) requires temperatures around 785°C[52]. These silicides are metallic[53,54] and their presence at the substrate-metal interface of superconducting devices might affect device performance. It is unlikely that these silicides are present at the S-M interface of our Ta resonator chips, since the maximum temperature used in the growth of these films is 500°C, and all depositions were completed in less than ten minutes. As demonstrated by Cheng et al.[43], the Ta-Si interfacial layer thickness can be controlled by modifying the annealing temperature or the annealing time. However, our STEM images indicate that the deposition temperature has no effect on the interfacial layer thickness (Supplementary Figures S2-S3). Further investigations on the effect of the deposition time at elevated temperatures will be needed to conclude if the Ta-Si interfacial layer can be controlled and if this layer negatively affects device performance.

The best $Q_{i,\,low}$ achieved in this work is $4.4\times10^6$ and corresponds to a resonator in a HF-treated chip. This value is around three times higher than the best α-Ta resonator $Q_{i,\,low}$ previously reported[29]. α-Ta transmon qubits with record $Q_i$ of $7.1\times10^6$ and $11.77\times10^6$ have also been reported by Place[18] and Wang[19], respectively. However, due to differences in device geometry a direct quality factor comparison between qubits and resonators is not accurate; nevertheless, the α-Ta films developed in this work are promising for the fabrication of high-quality qubits.

**Conclusion:**
This study demonstrates for the first-time the fabrication of high-quality α-Ta resonators directly grown on Si substrates over a large temperature processing window. The combination of spectroscopy and microscopy enables us to investigate the quality of the different resonator interfaces and material properties, and to establish correlations with device loss. Our analysis shows that surface oxides are the main contributor to resonator TLS loss, while intrinsic film properties such as crystallinity and resistivity do not seem to play any role at the current loss level. Our work opens the possibility for integrating α-Ta as a low-loss material in the fabrication of larger and more complex superconducting devices, as well as making it compatible with fabrication in industrial-grade facilities. The potential of this multidisciplinary approach to understand the material origin of loss sources paves the way to engineer high-performance superconducting devices, as well as other technologies that benefit from low-loss superconducting circuits.

**Methods:**
**Resonator fabrication.** Ta was deposited on 775 μm thick high-resistivity (> 3kΩcm) 300 mm Si (100) wafers. Prior to deposition, the Si was dip for 30 s in HF 2% to remove native $SiO_2$. The wafer was then loaded into physical vapor deposition system, where 100 nm of Ta were sputtered while the substrate was held at 400°C, 450°C and 500°C, respectively for each of the three different Ta films deposited. The device structures were patterned in a C and $SiO_2$ hard mask using ebeam lithography and mixture of $Cl_2/CH_2F_2$ reactive ion etching. Finally, the wafer was covered with protective resist and diced in individual chips, which were later cleaned using sonication for 5 min each in 50°C acetone and room temperature isopropanol, and finally cleaned by $O_2$ plasma for 2 minutes.

**Material characterization.** All material characterization measurements were performed at the Materials Characterization and Analysis Center at the IMEC microelectronics research institute.
TEM, STEM and EDS were performed in samples coated with spin-on carbon (SOC) layer. Lamellae with thickness of <50 nm were cut with focused ion beam (FIB) using Helios 450. TEM, STEM, and EDS measurements were performed on Titan Cubed Themis 300 STEM with a 200 kV source. For this study, TEM, EDS, atomic-resolution HAADF-STEM and ABF-STEM were used to investigate metal-air, substrate-air and substrate-metal interfaces.
ToF-SIMS measurements were performed using a TOFSIMS NCS instrument from ION-TOF GmbH. Negative ion profiles were measured in a dual beam configuration using a Bi+ (15keV) gun for analysis and a Cs+ (500eV) gun for sputtering, while positive ion profiles were measured in a dual beam configuration using a Bi+ (15keV) gun for analysis and a $O_2$+ (500eV) gun for sputtering.

XPS measurements were performed in Angle Integrated mode using a QUANTES instrument from Physical electronics. The measurements were performed using a monochromatized photon beam of 1486.6 eV. A 100 μm spot was used. Charge neutralization was used during this experiment.

Temperature dependence of the resistance of the Ta film, from 295 K down to 2 K, was performed in a PPMS (model ST-3T) using a Hall bar. Contacts to the films were established with 25 μm Al wires in a 4-point-measurement configuration. Current-voltage curves, from which the resistance of the films was extracted, were measured with three Keithley source-meter units (model 2636A) as current sources (one for each Ta film) and a Keithley nanovoltmeter (model 2182A).

**Resonator measurement setup**

Resonators are measured with a standard dilution-refrigerator setup. Input lines are thermalized with 20 dB attenuators at three different temperature stages (4 K, 0.1 K and 10mK). Output signal lines are thermalized with two isolators (LNF-ISC4_8A) with a total reverse isolation of ~ 40 dB and a 4-8 GHz band-pass filter (KBF-4/8-2S). The signal is amplified with a HEMT amplifier (LNF-LNC4_8C) at the 4 K stage and ultra-low noise amplifier (LNA-30-04000800-07-10P) at room temperature. In sample holder A superconducting qubit chips are glued and wire bonded to an aluminium sample holder. In sample holder B superconducting chips are pressed and wire bonded to an aluminium sample holder. Sample holder connectors are non-magnetic, and all samples are surrounded by a cryo-perm shield. Both sample holders are thermalized to the mixing chamber plate in a dilution refrigerator.

**Numerical Simulations of the Participation Ratios**

The electric field losses are dependent on and are proportional to the participation ratios of the dielectric regions in the devices. The participation ratio $p_i$ in a dielectric volume i is defined as the fraction of electrical energy contained in the dielectric with respect to the total electrical energy in the device, as follows[40]:

$$p_i = \frac{\epsilon_i \int_{V_i} |E_V|^2 dV}{\sum_i \epsilon_i \int_{V_i} |E_V|^2 dV}$$

where $\epsilon_i$ and $V_i$ correspond to the dielectric constant and volume of the $i^{th}$ dielectric respectively, and $|E_V|$ correspond to the absolute value of the electric field in the dielectric in an incremental volume $dV$. To calculate the participation ratio's $p_i$, we first estimate the electric fields by solving the Poisson's equation in the devices using a commercial electrostatic field simulation[55] and then subsequently incorporate the fields into the above equation. We invoke the following approximations for calculating the electric fields. With the geometry of the resonator being symmetric along its directions, we perform a two-dimensional electrostatic simulation, rather than a full 3-dimensional calculation. This further offers the advantage of reduced computational time and larger numerical accuracies for the simulations with an extremely refined mesh at interfaces in the device. Third, we only consider metal-air (M-A) and substrate-air (S-A) oxide interfaces in the simulations, while omitting the metal-substrate (M-S) interface, since we cannot experimentally distinguish losses from the M-S and from the substrate. Furthermore, uncertain parameters are the exact thickness of the oxide at the different interfaces and their dielectric constant. From TEM images, we estimate an average thickness 2.5 nm for the S-A interface in the three reference chips, M-A interfaces of 3.7 nm, 3.5 nm and 3.5 nm for the top part of the 400°C, 450°C and 500°C reference chips, and 6 nm, 6 and 6.5 for the sidewall, of the 400°C, 450°C and 500°C reference chips, respectively. The geometry of the structures used in the simulations is similar, although the simulated Si trench forms 90° wall with the bottom of the substrate instead of the angle visible in the STEM images. We also employ commonly chosen values for the dielectric constants i.e., 11.9, 3.9 and 25.0 for the silicon substrate, silicon dioxide and tantalum pentoxide, respectively. Based on the above technique, approximations, and parameters, we estimate the participation ratios in the capacitor regions for three reference resonators, shown in Supplementary Tables 2-4. For the HF-treated resonator simulation we scale the participation ratio of the M-A interface by the percentage of tantalum pentoxide left after the HF treatment compared to the reference resonators, shown in Supplementary Tables 5-7.

## Data availability
The data that supports the findings of this study is available from the corresponding authors upon reasonable request.

## Code availability
The codes are available upon reasonable request.

## Acknowledgments
The authors gratefully thank Paola Favia, Olivier Richard, Chris Drijbooms, Ilse Hoflijk, Thierry Conard, Céline Noël, Valentina Spampinato, Alexis Franquet for metrology support. This work was supported in part by the imec Industrial Affiliation Program on Quantum Computing. We also thanks Nathalie de Leon for useful discussions and comments about this work. This project leading to this application has received funding from the ECSEL Joint Undertaking (JU) under grant agreement No 101007322. The JU receives support from the European Union's Horizon 2020 research and innovation program and Germany, France, Belgium, Austria, Netherlands, Finland, Israel. (Please visit the project website www.matqu.eu for more information)

## Author information
**Affiliations**
**KU Leuven, Kasteelpark Arenberg 10, Leuven, B-3001, Belgium**
J. Verjauw, R. Acharya, J. Van Damme and K. De Greve.

**Imec, Kapeldreef 75, Leuven, B-3001, Belgium** D. P Lozano, M. Mongillo, X. Piao, S. Couet, D. Wan, Y. Canvel, A. M. Vadiraj, Ts. Ivanov, J. Verjauw, R. Acharya, J. Van Damme, F. A. Mohiyaddin, J. Jussot, P. P. Gowda, A. Pacco, B A. Potočnik, I. P. Radu, K. De Greve, B. Govoreanu and J. Swerts

**KU Leuven, Celetijnlaan 200D, Leuven, B-3001, Belgium** B. Raes and J. Van de Bondel.

## Author contributions
D. P Lozano and M. Mongillo planned the experiment. A. Potočnik. and M. Mongillo. designed the samples. D. P Lozano, X. Piao, S. Couet, J. Jussot, Y. Canvel, and D. Wan performed the resonator fabrication, with contributions from Ts. Ivanov. Resonator data was collected by M. Mongillo. and analysed by D. P Lozano, with contributions from A. M. Vadiraj, J. Verjauw, R. Acharya, and J. Van Damme. B. Raes. and J. Van de Vondel performed electrical characterization. F. A. Mohiyaddin. performed the participation ratio simulations. P. P. Gowda and A. Pacco provided the conditions for the HF treatment. D. P Lozano and A. Potočnik prepared the manuscript, with input from all authors. I. P. Radu, B. Govoreanu, K. De Greve and J. Swerts supervised and coordinated the project.

**Corresponding authors**
Correspondence to D. P. Lozano [Daniel.PerezLozano@imec.be]

# Supplementary information

## XRD characterization

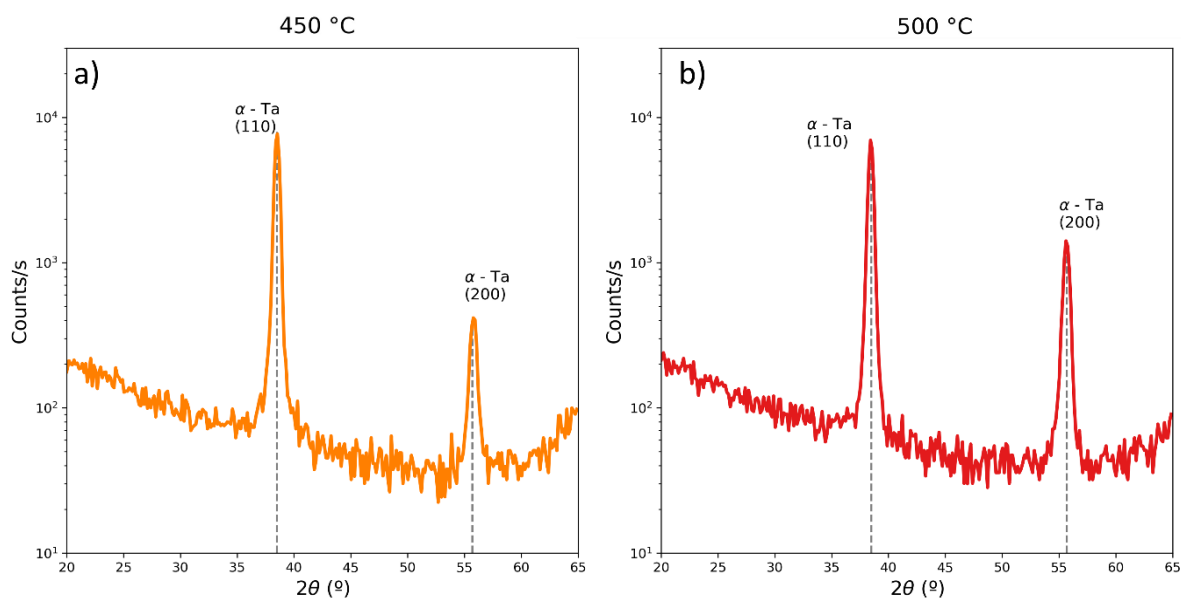

Supplementary Figure S1. XRD spectra of the other two films used for resonator fabrication. (a) XRD spectrum of the Ta film deposited at 450°C. (b) XRD spectrum of the Ta film deposited at 500°C

## TEM characterization

### 450°C reference α-Ta resonator

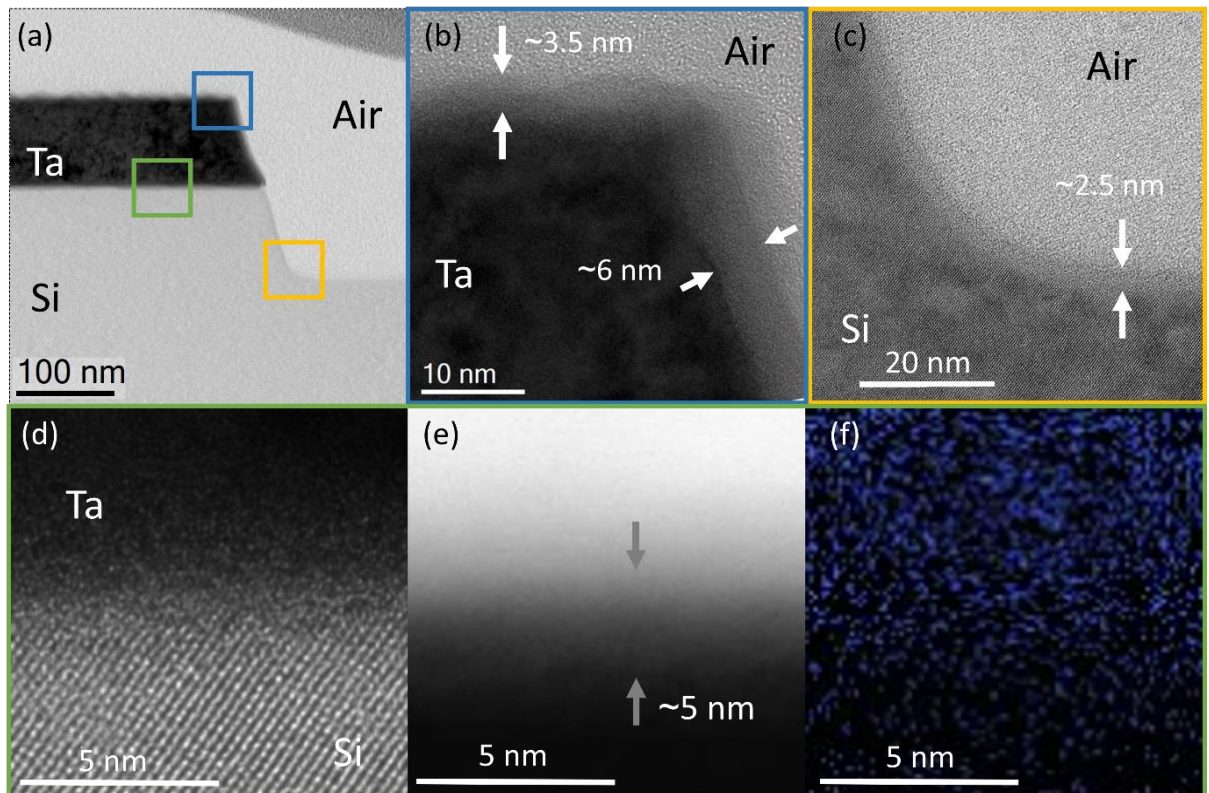

Supplementary Figure S2. STEM and EDS images of a α-Ta 450°C resonator cross section. (a) low magnification STEM cross section of Ta resonator. The coloured squares mark the areas where high magnification images were taken. (b) STEM cross section of the metal-air interface. Arrows indicate the tantalum oxide layer. (c) Annular bright-field-STEM image of the substrate-air interface. The arrows indicate the silicon oxide layer present. (d) STEM cross section of the substrate-metal

## 500°C reference α-Ta resonator

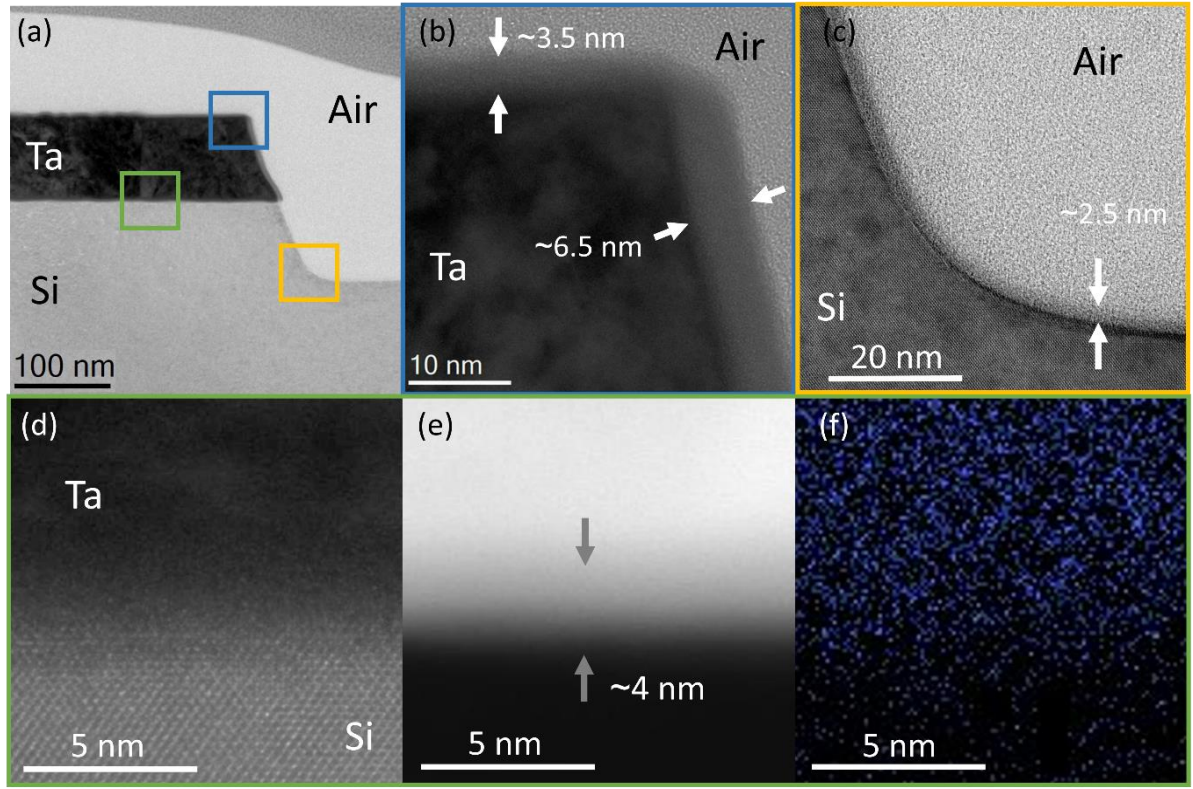

Supplementary Figure S3. STEM and EDS images of a 500°C α-Ta resonator cross section. (a) low magnification STEM cross section of Ta resonator. The coloured squares mark the areas where high magnification images were taken. (b) STEM cross section of the metal-air interface. Arrows indicate the tantalum oxide layer. (c) Annular bright-field-STEM image of the substrate-air interface. The arrows indicate the silicon oxide layer present. (d) STEM cross section of the substrate-metal interface. (e) High-angle annular dark-field-STEM cross section of the substrate-metal interface. The arrows indicate the extension of the Si-Ta interfacial layer. (f) Cross section oxygen EDS map of the substrate-metal interface.

## Resonator microwave characterization

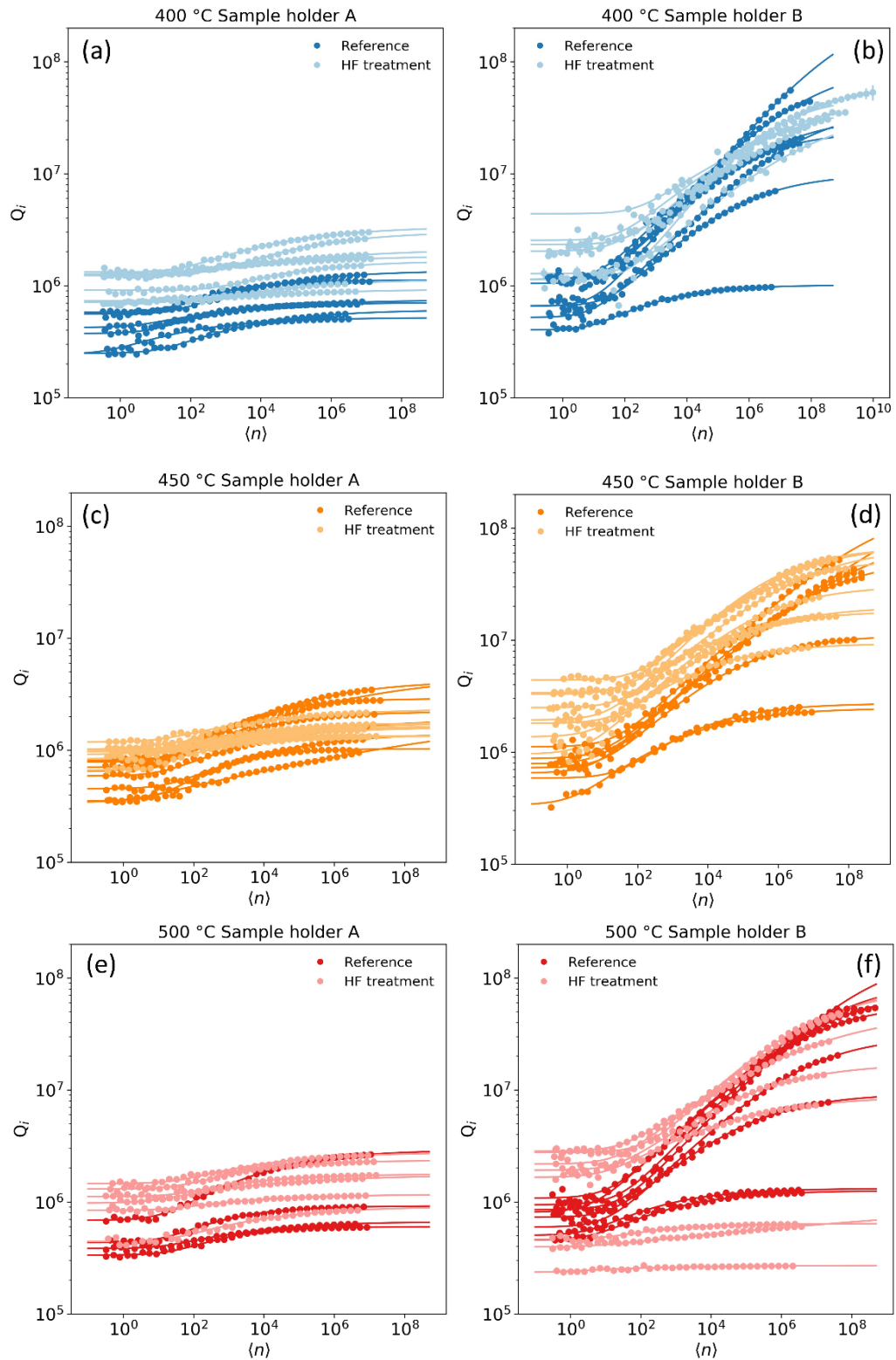

Supplementary Figure S4. Resonator quality factor measurements as a function of mean photon number for reference and HF-treated chips using sample holders A and B. (a-b) 400°C α-Ta films, (c-d) 450°C α-Ta films and (e-f) 500°C α-Ta films. Lines are fitting to Eq.1 in the main text.

## Resonator fitting summary

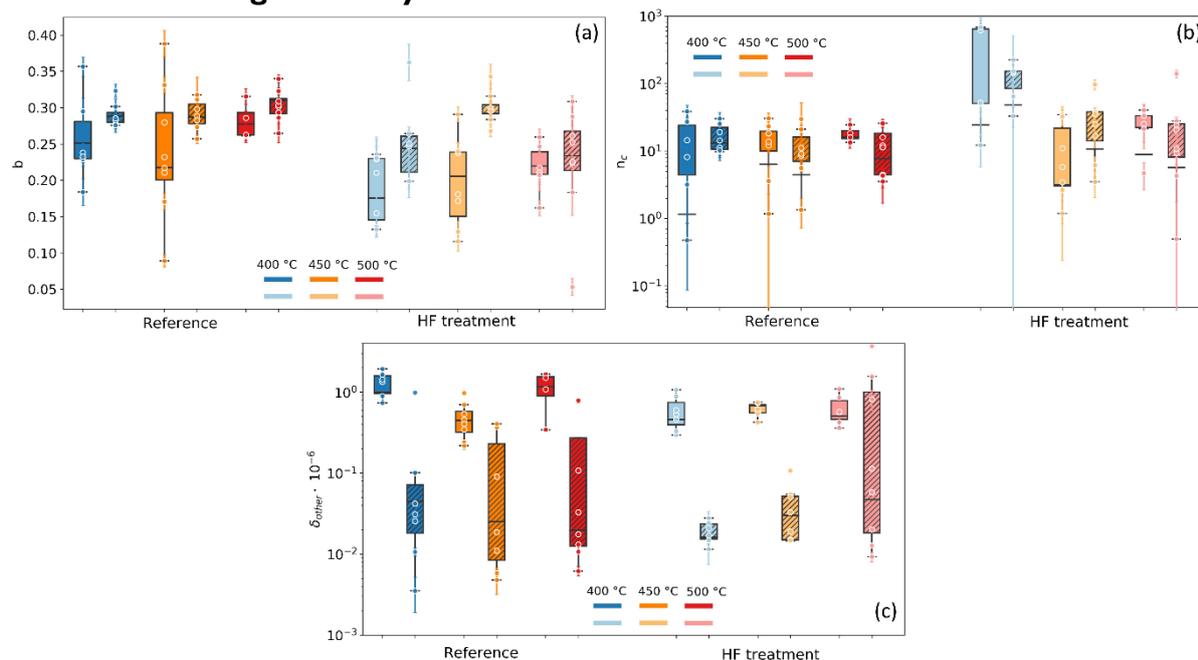

Supplementary Figure S5. TLS model fitting parameter summary. (a) Boxplot summary of b parameter for the three reference and HF-treated Ta resonators chips. (b) Boxplot summary of critical photon number for the three reference and HF-treated Ta resonators chips. (c) Boxplot summary of the power independent loss for the three reference and HF-treated Ta resonators chips. The boxes extend from the first quartile to the third quartile of the represented magnitude with a horizontal line representing the weighted mean. The whiskers extend from the box by 1.5 times the inter-quartile range. The dots indicate individual measurements. Boxes with no texture correspond to measurements taken using sample holder A, while boxes with a hatched pattern correspond to measurements taken using sample holder B

## XPS characterization

To determine the amount of tantalum oxide removed at the M-A interface after the HF treatment we performed an X-ray photo spectroscopy (XPS) study on reference and HF-treated Ta films deposited at 400°C, 450°C and 500°C. Four Ta peaks can be detected, as shown in Figure 5(a). The two peaks at low binding energy correspond to metallic tantalum orbitals $4f^{7/2}$ and $4f^{5/2}$, while the two peaks at higher binding energy correspond to the same orbitals but for tantalum pentoxide ($Ta_2O_5$). Note the strong reduction on the tantalum oxides peaks after the HF treatment and that there is no qualitative difference between Ta films deposited at the three temperatures.

The XPS spectra are best fit with three doublet components, revealing that there are two coexisting oxidation states besides Ta metal: $TaO_x$, and $Ta_2O_5$. Using this fitting model, the atomic percentages of tantalum and tantalum oxides on reference and HF-treated chips can be quantified. Results are presented in table supplementary table 1.

**Supplementary Table 1**. Atomic concentration table and calculated ratio. Columns indicate the percentage of metallic tantalum, tantalum suboxides and tantalum pentoxide out of the total of elements detected.

| Sample | Ta4f met.(%) | Ta4f subox.(%) | Ta4f pentox.(%) | (subox.+ pentox.) / Total Ta4f |
|---|---|---|---|---|
| **400°C Reference** | 3.72 | 1.27 | 20.05 | 0.85 |
| **450°C Reference** | 3.69 | 1.36 | 20.05 | 0.85 |
| **500°C Reference** | 3.74 | 1.32 | 20.23 | 0.85 |
| **400°C HF-treated** | 8.62 | 2.71 | 17.16 | 0.70 |
| **450°C HF-treated** | 7.78 | 2.69 | 16.82 | 0.71 |
| **500°C HF-treated** | 7.39 | 2.55 | 17.04 | 0.73 |

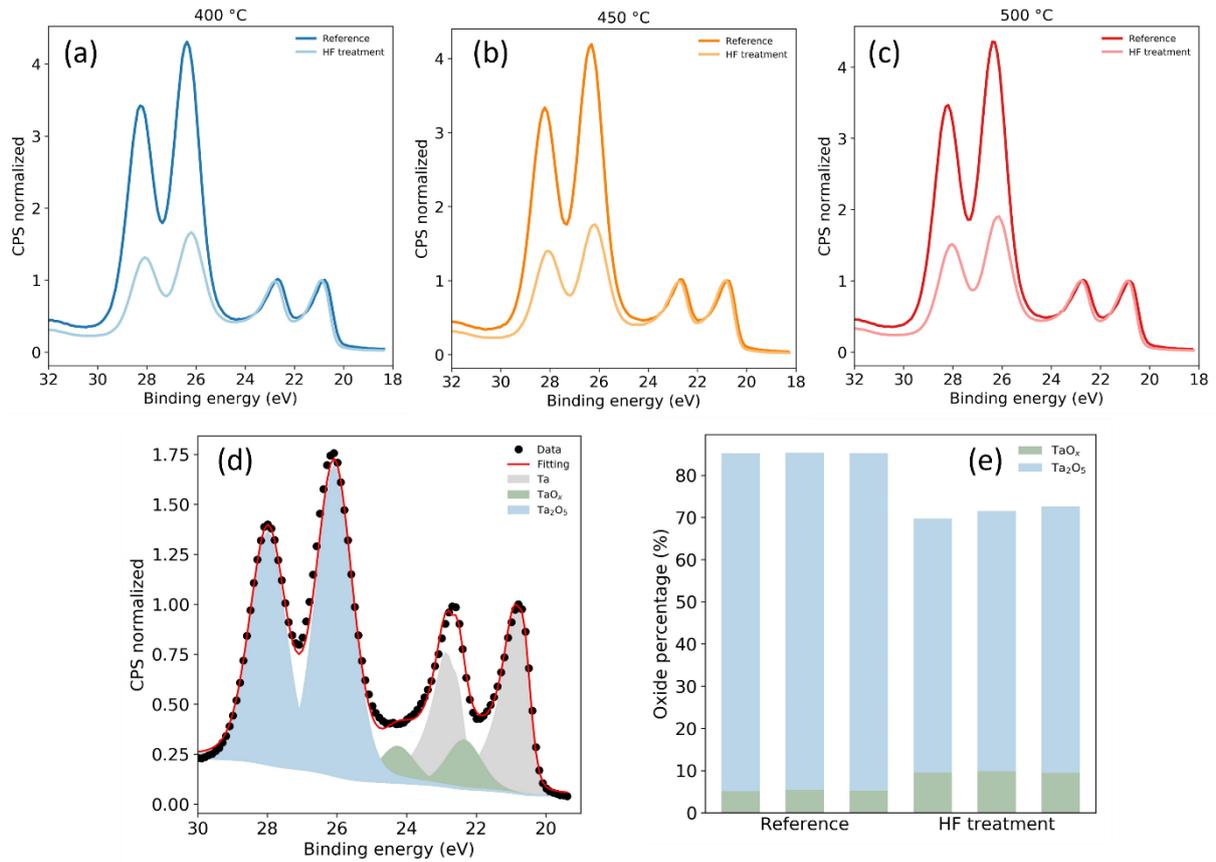

Supplementary Figure S6. XPS analysis of reference and HF-treated chips. (a-c) XPS spectra of a reference and HF-treated Ta chips fabricated using the 400°C, 450°C and 500°C Ta films. (d) Fitting of the XPS spectra of a HF-treated chip fabricated using the 400°C Ta film. Black dots are experimental data, red line is the fitting. The blue area represents the contribution to the fitting of tantalum pentoxide ($Ta_2O_5$), the grey area corresponds to the contribution of metallic Ta, and the green area the contribution from tantalum suboxides ($TaO_x$). (e) $Ta_2O_5$ and $TaO_x$ percentage respect to the total amount of Ta detected in the reference and HF-treated chips.

## Resonator loss simulations

**Supplementary Table 2**: Simulated participation ratios for the different dielectrics considered in the simulation for the reference 400°C Ta resonator chips.

| 400°C Reference | Participation ratio $F_i$ | tanδ | $F_i$tanδ |
|---|---|---|---|
| Silicon substrate | 0.911 | $1.3 \times 10^{-7}$ | $1.18 \times 10^{-7}$ |
| Air | 0.088 | 0 | 0 |
| Metal-Air | $1.87 \times 10^{-5}$ | 0.01 | $1.87 \times 10^{-7}$ |
| Substrate-Air | $3.7 \times 10^{-4}$ | $1.7 \times 10^{-3}$ | $6.28 \times 10^{-7}$ |
| Total loss | | | $9.34 \times 10^{-7}$ |

**Supplementary Table 3**: Simulated participation ratios for the different dielectrics considered in the simulation for the reference 450°C Ta resonator chips.

| 450°C Reference | Participation ratio $F_i$ | tanδ | $F_i$tanδ |
|---|---|---|---|
| Silicon substrate | 0.911 | $1.3 \times 10^{-7}$ | $1.18 \times 10^{-7}$ |
| Air | 0.088 | 0 | 0 |
| Metal-Air | $1.83 \times 10^{-5}$ | 0.01 | $1.83 \times 10^{-7}$ |
| Substrate-Air | $3.7 \times 10^{-4}$ | $1.7 \times 10^{-3}$ | $6.29 \times 10^{-7}$ |
| Total loss | | | $9.30 \times 10^{-7}$ |

**Supplementary Table 4**: Simulated participation ratios for the different dielectrics considered in the simulation for the reference 500°C Ta resonator chips.

| 500°C Reference | Participation ratio $F_i$ | $\tan\delta$ | $F_i\tan\delta$ |
|---|---|---|---|
| Silicon substrate | 0.911 | $1.3\times10^{-7}$ | $1.18\times10^{-7}$ |
| Air | 0.088 | 0 | 0 |
| Metal-Air | $1.95\times10^{-5}$ | 0.01 | $1.95\times10^{-7}$ |
| Substrate-Air | $3.94\times10^{-4}$ | $1.7\times10^{-3}$ | $6.69\times10^{-7}$ |
| Total loss | | | $9.83\times10^{-7}$ |

**Supplementary Table 5**: Simulated participation ratios for the different dielectrics considered in the simulation for the HF-treated 400°C Ta resonator chips.

| 400°C HF treated | Participation ratio $F_i$ | $\tan\delta$ | $F_i\tan\delta$ |
|---|---|---|---|
| Silicon substrate | 0.911 | $1.3\times10^{-7}$ | $1.18\times10^{-7}$ |
| Air | 0.088 | 0 | 0 |
| Metal-Air | $1.53\times10^{-5}$ | 0.01 | $1.53\times10^{-7}$ |
| Substrate-Air | 0 | 0 | 0 |
| Total loss | | | $2.72\times10^{-7}$ |

**Supplementary Table 6**: Simulated participation ratios for the different dielectrics considered in the simulation for the HF-treated 450°C Ta resonator chips.

| 450°C HF-treated | Participation ratio $F_i$ | $\tan\delta$ | $F_i\tan\delta$ |
|---|---|---|---|
| Silicon substrate | 0.911 | $1.3\times10^{-7}$ | $1.18\times10^{-7}$ |
| Air | 0.088 | 0 | 0 |
| Metal-Air | $1.53\times10^{-5}$ | 0.01 | $1.53\times10^{-7}$ |
| Substrate-Air | 0 | 0 | 0 |
| Total loss | | | $2.72\times10^{-7}$ |

**Supplementary Table 7**: Simulated participation ratios for the different dielectrics considered in the simulation for the HF-treated 500°C Ta resonator chips.

| 500°C HF-treated | Participation ratio $F_i$ | $\tan\delta$ | $F_i\tan\delta$ |
|---|---|---|---|
| Silicon substrate | 0.911 | $1.3\times10^{-7}$ | $1.18\times10^{-7}$ |
| Air | 0.088 | 0 | 0 |
| Metal-Air | $1.66\times10^{-5}$ | 0.01 | $1.66\times10^{-7}$ |
| Substrate-Air | 0 | 0 | 0 |
| Total loss | | | $2.85\times10^{-7}$ |